\documentclass[RNAAS]{aastex62}

\begin{document}

\title{Detection of a glitch in the pulsar J1709$-$4429}

\correspondingauthor{Marcus E. Lower}

\author[0000-0001-9208-0009]{Marcus E. Lower}
\affiliation{Centre for Astrophysics and Supercomputing, Swinburne University of Technology, VIC 3122, Australia}
\email{mlower@swin.edu.au}

\author[0000-0003-1110-0712]{Chris Flynn}
\affiliation{Centre for Astrophysics and Supercomputing, Swinburne University of Technology, VIC 3122, Australia}
\affiliation{ARC Centre of Excellence for All-sky Astrophysics (CAASTRO)}

\author[0000-0003-3294-3081]{Matthew Bailes}
\affiliation{Centre for Astrophysics and Supercomputing, Swinburne University of Technology, VIC 3122, Australia}

\author{Ewan D. Barr}
\affiliation{Max-Plank-Institute f\"{u}r Radioastronomie, Auf dem H\"{u}gel 69, D-53121 Bonn, Germany}

\author{Timothy Bateman}
\affiliation{Sydney Institute for Astronomy, School of Physics, A28, University of Sydney, NSW 2006, Australia}

\author{Shivani Bhandari}
\affiliation{Centre for Astrophysics and Supercomputing, Swinburne University of Technology, VIC 3122, Australia}
\affiliation{ARC Centre of Excellence for All-sky Astrophysics (CAASTRO)}
\affiliation{ATNF, CSIRO Astronomy and Space Science, PO Box 76, Epping, NSW 1710, Australia}

\author{Manisha Caleb}
\affiliation{ARC Centre of Excellence for All-sky Astrophysics (CAASTRO)}
\affiliation{Research School of Astronomy and Astrophysics, Australian National University, Canberra, ACT 2611, Australia}
\affiliation{Jodrell Bank Centre for Astrophysics, School of Physics and Astronomy, The University of Manchester, Manchester M13 9PL, UK}

\author{Duncan Campbell-Wilson}
\affiliation{Sydney Institute for Astronomy, School of Physics, A28, University of Sydney, NSW 2006, Australia}

\author{Cherie Day}
\affiliation{Centre for Astrophysics and Supercomputing, Swinburne University of Technology, VIC 3122, Australia}

\author[0000-0001-9434-3837]{Adam Deller}
\affiliation{Centre for Astrophysics and Supercomputing, Swinburne University of Technology, VIC 3122, Australia}
\affiliation{ARC Centre of Excellence for All-sky Astrophysics (CAASTRO)}

\author{Wael Farah}
\affiliation{Centre for Astrophysics and Supercomputing, Swinburne University of Technology, VIC 3122, Australia}

\author{Anne J. Green}
\affiliation{Sydney Institute for Astronomy, School of Physics, A28, University of Sydney, NSW 2006, Australia}

\author{Vivek Gupta}
\affiliation{Centre for Astrophysics and Supercomputing, Swinburne University of Technology, VIC 3122, Australia}

\author[0000-0002-3205-8288]{Richard W. Hunstead}
\affiliation{Sydney Institute for Astronomy, School of Physics, A28, University of Sydney, NSW 2006, Australia}

\author[0000-0002-0996-3001]{Andrew Jameson}
\affiliation{Centre for Astrophysics and Supercomputing, Swinburne University of Technology, VIC 3122, Australia}
\affiliation{ARC Centre of Excellence for All-sky Astrophysics (CAASTRO)}

\author[0000-0002-6658-2811]{Fabian Jankowski}
\affiliation{Centre for Astrophysics and Supercomputing, Swinburne University of Technology, VIC 3122, Australia}
\affiliation{ARC Centre of Excellence for All-sky Astrophysics (CAASTRO)}
\affiliation{Jodrell Bank Centre for Astrophysics, School of Physics and Astronomy, The University of Manchester, Manchester M13 9PL, UK}

\author[0000-0002-4553-655X]{Evan F. Keane}
\affiliation{ARC Centre of Excellence for All-sky Astrophysics (CAASTRO)}
\affiliation{SKA Organization, Jodrell Bank Observatory, Cheshire SK11 9DL, UK}

\author[0000-0001-9518-9819]{Vivek Venkatraman Krishnan}
\affiliation{Centre for Astrophysics and Supercomputing, Swinburne University of Technology, VIC 3122, Australia}
\affiliation{ARC Centre of Excellence for All-sky Astrophysics (CAASTRO)}

\author[0000-0003-0289-0732]{Stefan Os{\l}owski}
\affiliation{Centre for Astrophysics and Supercomputing, Swinburne University of Technology, VIC 3122, Australia}

\author{Aditya Parthasarathy}
\affiliation{Centre for Astrophysics and Supercomputing, Swinburne University of Technology, VIC 3122, Australia}
\affiliation{ARC Centre of Excellence for All-sky Astrophysics (CAASTRO)}

\author{Kathryn Plant}
\affiliation{Centre for Astrophysics and Supercomputing, Swinburne University of Technology, VIC 3122, Australia}
\affiliation{Cahill Centre for Astronomy and Astrophysics, MC 249-17, California Institute of Technology, Pasadena, CA 91125, USA}

\author[0000-0003-2783-1608]{Danny C. Price}
\affiliation{Centre for Astrophysics and Supercomputing, Swinburne University of Technology, VIC 3122, Australia}

\author{Vikram Ravi}
\affiliation{Cahill Centre for Astronomy and Astrophysics, MC 249-17, California Institute of Technology, Pasadena, CA 91125, USA}

\author[0000-0002-7285-6348]{Ryan M. Shannon}
\affiliation{Centre for Astrophysics and Supercomputing, Swinburne University of Technology, VIC 3122, Australia}

\author{David Temby}
\affiliation{Sydney Institute for Astronomy, School of Physics, A28, University of Sydney, NSW 2006, Australia}

\author{Glen Torr}
\affiliation{Sydney Institute for Astronomy, School of Physics, A28, University of Sydney, NSW 2006, Australia}

\author{Glenn Urquhart}
\affiliation{Sydney Institute for Astronomy, School of Physics, A28, University of Sydney, NSW 2006, Australia}

\keywords{stars: neutron --- 
pulsars: individual (J1709$-$4429)}

\section{Introduction}
\label{sec:intro}

Pulsar glitches are thought to result from either quakes in the neutron star crust \citep{baym69}, or by a transfer of angular momentum between the superfluid interior and the outer crust \citep{anderson75}. The event manifests as a sudden increase in the observed spin period and spin-down of the pulsar, which can be followed by a recovery phase where the period exponentially returns to its pre-glitch evolution.

We report here the detection of a glitch event in the pulsar J1709$-$4429 (also known as B1706$-$44) during regular monitoring observations with the Molonglo Observatory Synthesis Telescope (MOST). MOST is an aperture synthesis radio telescope located $40$\,km East of Canberra, Australia, operating in the 820--850\,MHz frequency range. The UTMOST backend upgrade to the MOST \citep{bailes17} has enabled study of the dynamic radio sky on millisecond timescales, and is well suited to pulsar timing, pulsar searches, observing single pulses from pulsars and discoveries of Fast Radio Bursts (FRBs) \citep{caleb17, farah18}. The glitch was found during timing operations, in which we regularly observe over 400 pulsars with up to daily cadence, while commensally searching for Rotating Radio Transients, pulsars, and FRBs.

\section{Glitch parameters}
\label{sec:params}

J1709$-$4429 is a bright (7.3\,mJy at 1400\,MHz) pulsar with a period of 0.102459\,s and a dispersion measure of 75.7\,pc\,cm$^{-3}$, for which 90 timing measurements have been made at UTMOST since May 2015. We constrain the epoch at which the glitch occurred ($t_{\mathrm{g}}$) to MJD $58178 \pm 6$. This is the median MJD between observations made on 02-23-2018 and 03-07-2018 UTC. The uncertainty in the glitch epoch is half the difference in time between the last pre-glitch and first post-glitch observations.
As the glitch epoch is relatively unconstrained, an unphysical jump in the pulsar phase of $\Delta\phi_{\mathrm{g}} = 3.39\pm0.01$ is required to achieve phase-connected timing residuals.

Using the \texttt{\sc Tempo2} \citep{hobbs06, edwards06} and \texttt{\sc TempoNest} \citep{lentati14} pulsar timing packages, we estimate the instantaneous change in spin frequency, spin frequency derivative and spin frequency second derivative to be $\Delta\nu = (516.07 \pm 3.6) \times10^{-9}$ Hz, $\Delta\dot{\nu} = (-6.46 \pm 0.11) \times10^{-14}$\,s$^{-2}$ and $\Delta\ddot{\nu} = (-22.42 \pm 1.8) \times10^{-22}$\,s$^{-3}$ respectively.

\begin{figure}[t!]
\begin{center}
\includegraphics[scale=0.6,angle=0]{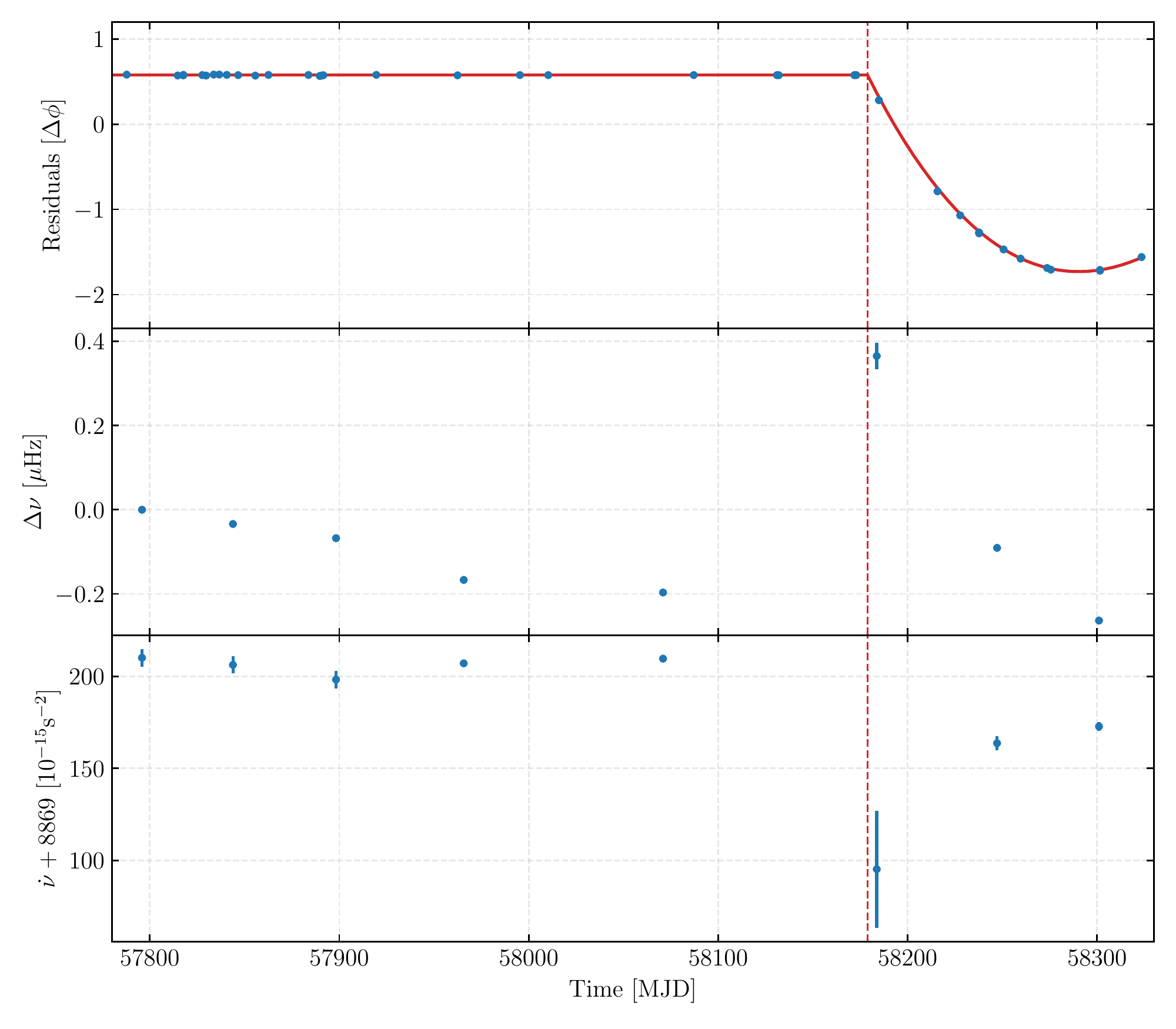}
\caption{Timing phase residuals for J1709-4429 in blue with the best-fit glitch model indicated in red (top), in addition to the frequency residuals from subtracting the pre-glitch slope (middle) and the change in $\dot{\nu}$ with time (bottom). The median glitch epoch is indicated by the dashed vertical line.}

\label{fig:fit}
\end{center}
\end{figure}

The upper panel of Figure~\ref{fig:fit} shows the timing residuals for J1709-4429 prior to fitting the recovered glitch parameters (solid blue points) and the best-fit glitch model (red line). The lower two panels show the evolution of $\Delta\nu$ and $\dot{\nu}$ with time. 

Four previous glitches in J1709$-$4429 are recorded in the \href{https://www.atnf.csiro.au/people/pulsar/psrcat/glitchTbl.html}{Australian National Telescope Facility} and Jodrell Bank glitch catalogues \citep{jb_glitchcat}. With a fractional size of $\Delta\nu/\nu = 52.4 \pm 0.1 \times10^{-9}$, the glitch reported here is by far the smallest known for this pulsar, attesting to the efficacy of glitch searches with high cadence using UTMOST.
Continued observations of J1709$-$4429 are being undertaken with UTMOST. We encourage monitoring of this pulsar by other timing programs.

\acknowledgments

The Molonglo Observatory is owned and operated by the University of Sydney. Major support for the UTMOST project has been provided by Swinburne University of Technology. 
We acknowledge the Australian Research Council grants CE110001020 (CAASTRO) and the Laureate Fellowship 
FL150100148.
This work made use of the gSTAR and OzStar national HPC facilities.

\bibliographystyle{aasjournal}
\bibliography{glitch}

\end{document}